\documentclass[a4paper,fleqn,usenatbib]{mnras}

\usepackage{newtxtext,newtxmath}

\usepackage[T1]{fontenc}
\usepackage{ae,aecompl}

\usepackage{graphicx}	
\usepackage{amsmath}	
\usepackage{amssymb}	

\usepackage{comment}

\usepackage{color}
\usepackage{ulem}

\title[Age and Metallicity Dependence of RC Magnitudes]{The Age and Metallicity Dependence of the Near-Infrared Magnitudes of Red Clump Stars}

\author[H. Onozato et al.]{
Hiroki Onozato,$^{1, 2}$\thanks{E-mail: onozato@nhao.jp}
Yoshifusa Ita,$^{1}$
Yoshikazu Nakada$^{3}$
and Shogo Nishiyama$^{4}$
\\
$^{1}$Astronomical Institute, Graduate School of Science, Tohoku University, 6-3, Aramaki Aoba, Aoba-ku, Sendai, Miyagi 980-8578, Japan\\
$^{2}$Nishi-Harima Astronomical Observatory, Center for Astronomy, Institute of Natural and Environmental Sciences, University of Hyogo,\\
407-2, Nishigaichi, Sayo-cho, Sayo-gun, Hyogo 679-5313, Japan\\
$^{3}$Kiso Observatory, Institute of Astronomy, School of Science, The University of Tokyo,\\
10762-30 Mitake, Kiso-machi, Kiso-gun, Nagano 397-0101, Japan\\
$^{4}$Miyagi University of Education, 149, Aramaki Aoba, Aoba-ku, Sendai, Miyagi 980-0845, Japan
}

\date{Accepted XXX. Received YYY; in original form ZZZ}

\pubyear{2018}

\begin{document}
\label{firstpage}
\pagerange{\pageref{firstpage}--\pageref{lastpage}}
\maketitle

\begin{abstract}
Red clump (RC) stars are widely used as an excellent standard candle. To make them even better, it is important to know the dependence of their absolute magnitudes on age and metallicity. We observed star clusters in the Large Magellanic Cloud to fill age and metallicity parameter space, which previous work has not observationally studied. We obtained the empirical relations of the age and metallicity dependence of absolute magnitudes $M_{J}$, $M_{H}$, and $M_{K_{S}}$, and colours $J - H$, $J - K_{S}$, and $H - K_{S}$ of RC stars, although the coefficients have large errors. Mean near-infrared magnitudes of the RC stars in the clusters show relatively strong dependence on age for young RC stars. The $J - K_{S}$ and $H - K_{S}$ colours show the nearly constant values of $0.528 \pm 0.015$ and $0.047 \pm 0.011$, respectively, at least within the ages of 1.1--3.2~Gyr and [Fe/H] of $-0.90$ to $-0.40$~dex. We also confirmed that the population effects of observational data are in good agreement with the model prediction.
\end{abstract}




\section{INTRODUCTION}
Red clump (RC) stars are low mass stars in the stage of the core helium-burning phase. They are easily recognised in the Heltzsprung-Russell diagram, because they have similar luminosity and effective temperature, and they are numerous. They were first recognised in the colour-magnitude diagrams of intermediate-age clusters \citep{C1970}. After the work by \citet{PS1998} using \textit{Hipparcos} parallaxes to determine the absolute magnitudes of RC stars, they gained much attention and have been widely used as an ideal standard candle to investigate Galactic structure \citep[e.g.,][]{MZ2010, NUG2010} and interstellar extinction \citep[e.g.,][]{NNK2006, NTH2009, NGF2013}.

To make them an even better standard candle, it is important to understand the dependence of their absolute magnitudes on age and metallicity (population effects). From the theoretical side, \citet{GS2001} investigated the population effects on the $V$- and $I$-band absolute magnitudes, and \citet{SG2002} studied the effects in the $K$-band. They suggested that the longer the wavelength becomes, the smaller absolute magnitude depends on metallicity, and it does not depend much on metallicity in near-infrared (NIR) wavelengths. Meanwhile, they predict that age dependence of the absolute magnitude is not simple such that there is only weak age dependence for RC stars older than 2~Gyr, but there is strong age dependence for RC stars younger than 2~Gyr. Many observational studies have confirmed small metallicity dependence of NIR magnitudes among RC stars in the solar neighbourhood \citep{A2000, G2008, LJP2012} and Milky Way star clusters \citep{GS2002, PS2003, vHG2007}. On the other hand, age dependence has not been investigated extensively because it is difficult to know the ages of RC stars. So far, the studies are limited to the work using Milky Way star clusters \citep{GS2002, PS2003, vHG2007} or age that derived from asteroseismology \citep{CCZ2017}, and no clear age dependence has been observed for younger RC stars. This is mainly due to the small number of samples for the former case, and the relatively large uncertainties in age for the latter case. Therefore, the number of samples with both age and metallicity information is quite limited, and the parameter space is not covered enough up until now.

Hence, we use star clusters in the Large Magellanic Cloud (LMC) to fill the parameter space, which previous research has not observationally studied. Star clusters in the LMC have different age and metallicity from Milky Way star clusters. Star clusters in the LMC are more metal-poor than that in the Milky Way. Therefore, we can expand the parameter space to the more metal-poor and younger range.  Moreover, the LMC is distant enough to be considered as the stars in the galaxy are at the same distance, so the uncertainty of the distance does not much affect the determination of absolute magnitudes of RC stars. In addition to these advantages, there are many young star clusters in the LMC, so we can investigate the age dependence of young RC stars where large age dependence is predicted from theoretical models. RC stars in the LMC star clusters have not been studied because they are too faint to determine their mean magnitude reliably in past NIR surveys such as two micron all sky survey (2MASS). Hence, we conducted NIR observations of the clusters using Infrared Survey Facility (IRSF). Long exposure time of our observations makes it possible to determine the mean RC magnitude in the LMC clusters.

It is important to know population effects on RC luminosity in terms of stellar evolution as well as a standard candle. The difference of age dependence between young and old RC stars comes from whether they experience helium flash or not. Because the age of an RC star corresponds to their mass, investigation of age dependence leads to better understanding of the maximum mass of stars that will experience helium flash \citep{G2016}.

In section \ref{sec:Data}, we show the data and methods to determine NIR magnitudes of RC stars. Results and comparison with previous work are presented in section \ref{sec:Results}. Section \ref{sec:Conclusions} is our conclusions.

\section{The Data}\label{sec:Data}
\subsection{Sample selection}\label{sec:sample}
When we investigate the metallicity and age dependence of absolute magnitudes of RC stars in clusters, it is important that metallicities and ages of the clusters have been determined using the same techniques. We can avoid systematic errors by using uniform samples. As a catalogue that satisfies this condition, we used the LMC star cluster catalogue compiled by \citet{PGC2016}. Palma's catalogue lists 277 LMC star clusters.

\begin{figure*}
    \includegraphics{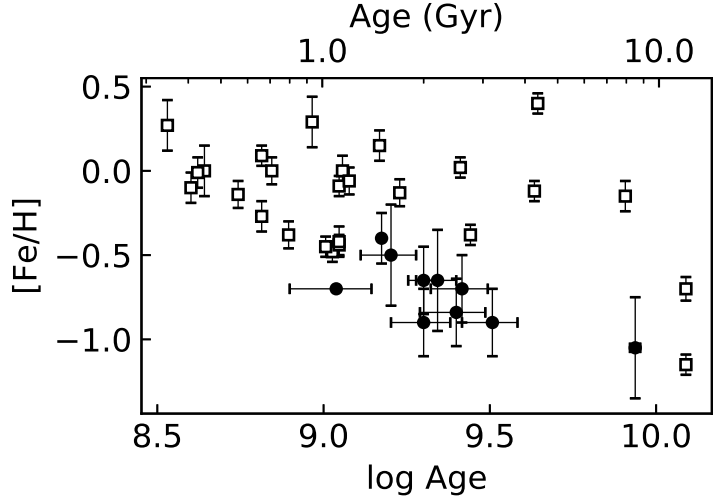}
    \caption{Age and metallicity distribution of our target clusters (filled circles) and clusters observed by \citet[open squares]{vHG2007}. }
    \label{fig:Age_and_Metallicity}
\end{figure*}

From the catalogue, we first selected clusters that had both age and metallicity information. Then, we chose clusters whose radii are larger than $0\farcm65$ (9.5~pc at 50~kpc) as our target clusters because we cannot detect the excess of RC stars for small clusters. We excluded clusters that are in the bar region from the target since these clusters are heavily contaminated by field stars. The number of clusters at each selection process is shown in Table~\ref{tab:number}. Finally, 15~clusters were selected to be observed by ourselves. In the 15~clusters, we have detected a clear RC peak in its luminosity function for 10~clusters (see below for more detail). Figure~\ref{fig:Age_and_Metallicity} shows age and metallicity distribution of our target clusters, compared to clusters observed by \citet{vHG2007}. Most of our target clusters in the LMC ranges from 1 to 3~Gyr in age and from -1.0 to -0.4 in [Fe/H], where \citet{vHG2007} have few samples. In this age range, a strong age dependence of the RC absolute magnitudes is predicted. Our target clusters have lower metallicity than the clusters used in \citet{vHG2007}. In the very low metallicity region, theoretical models predict relatively strong metallicity dependence of absolute magnitudes. Our sample clusters allow us to investigate the dependence of the RC magnitudes at lower metallicity range than in  our Galaxy.

\begin{table*}
\caption{The number of star clusters at each selection process}
\label{tab:number}
\begin{tabular}{lc}
\hline
Selection process & number of clusters\\
\hline
clusters in \citet{PGC2016} & 277\\
clusters with age and metallicity &  176\\
clusters larger than $0\farcm65$ & 26\\
clusters in outer region (observed) & 15\\
clusters that have significant RC excess and can be fitted by equation~(\ref{eq:fitting}) & 10\\
\hline
\end{tabular}
\end{table*}

\subsection{Observation}\label{sec:Observation}
Observations were performed using the SIRIUS camera equipped on the IRSF 1.4-m telescope \citep{NNN1999, NNN2003} in the South African Astronomical Observatory in 2017 November and December. SIRIUS can collect $JHK_{S}$-band images simultaneously with a $7\farcm7 \times 7\farcm7$ field of view with a pixel scale of $0\farcs45$ pixel$^{-1}$. The seeing size was typically 1.5~arcsec and sometimes reached to 0.9~arcsec. The exposure time of each image was 20 s and 25 images were taken in each dithering set. The number of observation sets were from 35 to 41. Observed clusters and their observational information are listed in Table \ref{tab:observed}.

\begin{table*}
\caption{List of observed star clusters}
\label{tab:observed}
\begin{tabular}{lccccc}
\hline
Cluster name & RA (J2000) & Dec (J2000) & Observation Date & Number of combined images & total exposure time (sec)\\
\hline
KMHK 21 & 04$^{\mathrm{h}}$ 37$^{\mathrm{m}}$ 52$^{\mathrm{s}}$ & -69\degr 01\arcmin 42\arcsec & 2017 Nov 16 & 925 & 18500\\
KMHK 337 & 04$^{\mathrm{h}}$ 57$^{\mathrm{m}}$ 34$^{\mathrm{s}}$ & -65\degr 16\arcmin 00\arcsec & 2017 Nov 22 & 1025 & 20500\\
ESO 85-72 & 05$^{\mathrm{h}}$ 20$^{\mathrm{m}}$ 05$^{\mathrm{s}}$ & -63\degr 28\arcmin 49\arcsec & 2017 Nov 11 &900 & 19000\\
NGC 1997 & 05$^{\mathrm{h}}$ 30$^{\mathrm{m}}$ 34$^{\mathrm{s}}$ & -63\degr 12\arcmin 12\arcsec & 2017 Nov 10 & 800 & 16000\\
IC 2140 & 05$^{\mathrm{h}}$ 33$^{\mathrm{m}}$ 21$^{\mathrm{s}}$ & -75\degr 22\arcmin 35\arcsec & 2017 Dec 3 & 925 & 18500\\
KMHK 1281 & 05$^{\mathrm{h}}$ 43$^{\mathrm{m}}$ 20$^{\mathrm{s}}$ & -66\degr 15\arcmin 44\arcsec & 2017 Dec 12 & 975 & 19500\\
NGC 2161 & 05$^{\mathrm{h}}$ 55$^{\mathrm{m}}$ 42$^{\mathrm{s}}$ & -74\degr 21\arcmin 14\arcsec & 2017 Nov 27 & 1000 & 20000\\
NGC 2155 & 05$^{\mathrm{h}}$ 58$^{\mathrm{m}}$ 33$^{\mathrm{s}}$ & -65\degr 28\arcmin 37\arcsec & 2017 Nov 30 & 1000 & 20000\\
ESO 121-3 & 06$^{\mathrm{h}}$ 02$^{\mathrm{m}}$ 02$^{\mathrm{s}}$ & -60\degr 31\arcmin 24\arcsec & 2017 Nov 9 & 875 & 17500\\
NGC 2213 & 06$^{\mathrm{h}}$ 10$^{\mathrm{m}}$ 42$^{\mathrm{s}}$ & -71\degr 31\arcmin 44\arcsec & 2017 Dec 11 & 975 & 19500\\
\hline
\end{tabular}
\end{table*}

\subsection{Data reduction}\label{sec:reduction}
\begin{figure*}
    \includegraphics{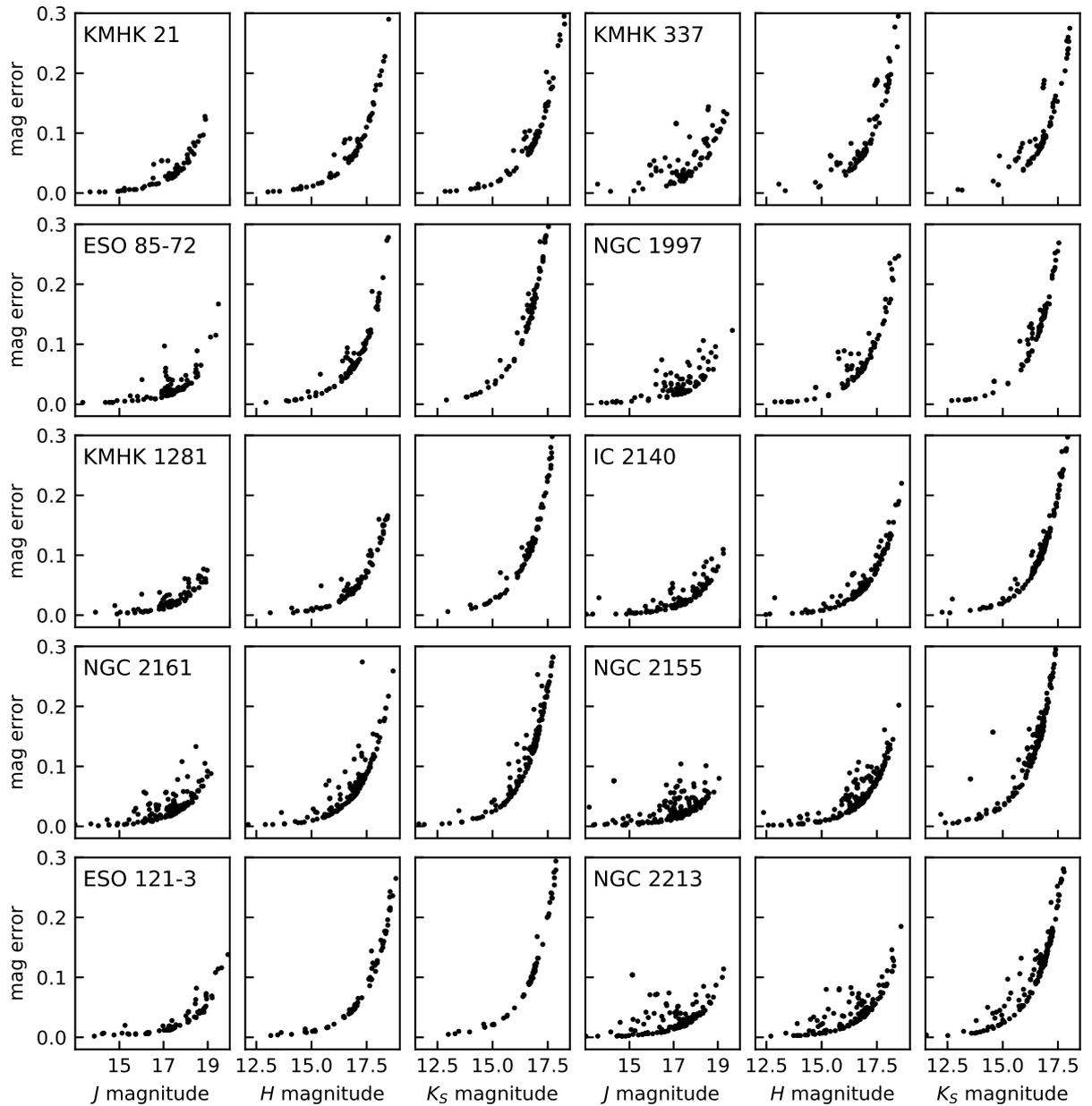}
    \caption{Photometric errors versus magnitudes for our target clusters.}
    \label{fig:Mag_error}
\end{figure*}

We reduced the obtained images by standard data reduction process, which consisted of dark subtraction, flat-fielding, sky subtraction, and dithered-image-combining. We used pyIRSF pipeline software\footnote{https://sourceforge.net/projects/irsfsoftware/} for this reduction. Point spread function fitting photometry is performed  with \textsc{iraf/daophot} package. We used the 2MASS point source catalogue \citep{SCS2006} to convert apparent magnitudes to calibrated apparent magnitudes. We calculated the weighted mean difference between instrumental magnitude and 2MASS magnitude in the field of view to decide zero points. Photometric errors against $JHK_{S}$ magnitudes for stars in our target clusters are plotted in Figure \ref{fig:Mag_error}. These errors were calculated by the allstar task in \textsc{iraf/daophot} package. Typical error of $J$-band is 0.04~mag at 18.0~mag, $H$-band is 0.08~mag at 17.5~mag, and $K_{S}$ band is 0.20~mag at 17.5~mag.

\begin{figure*}
    \includegraphics{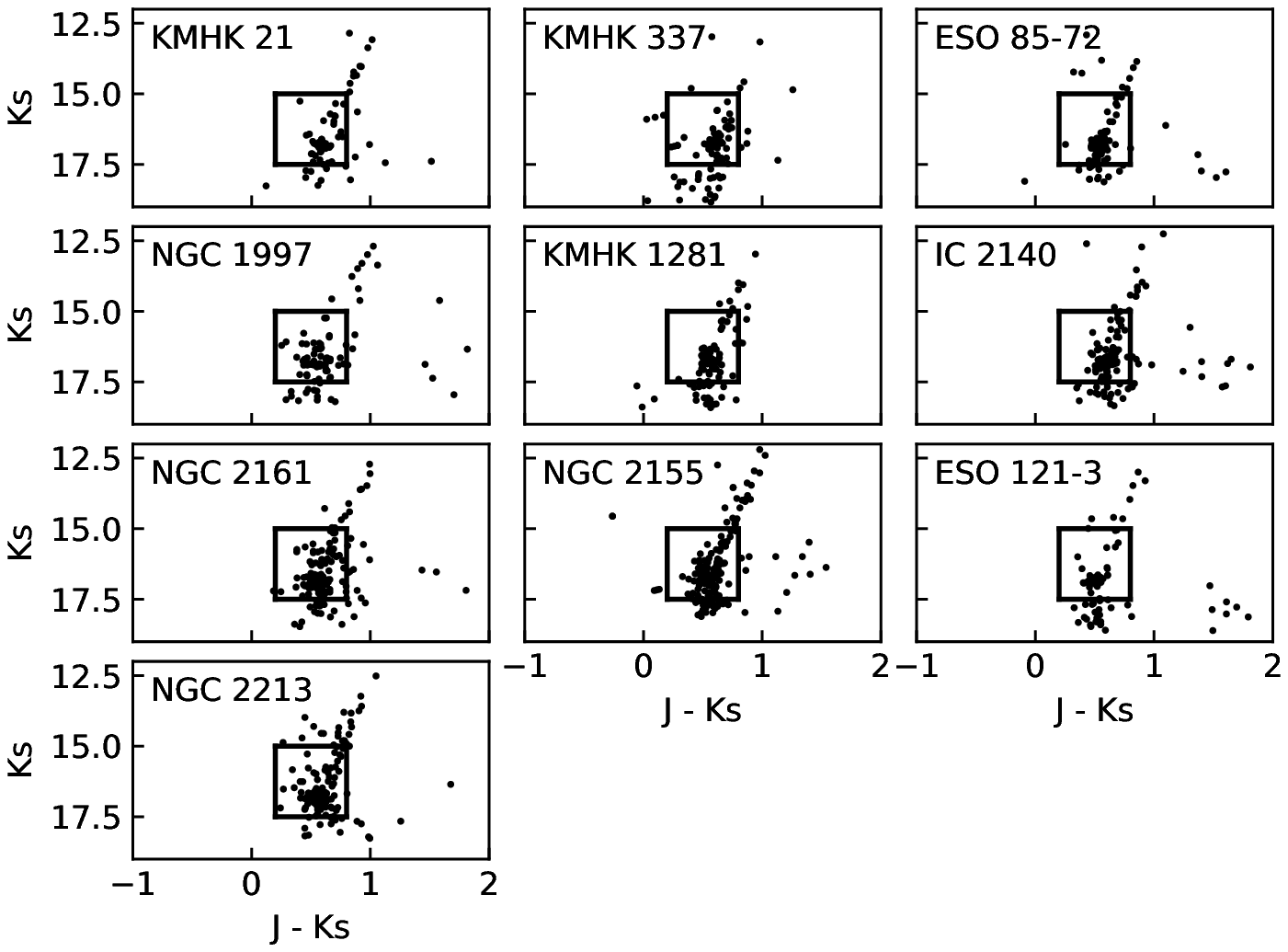}
    \caption{$J - K_{S}$ versus $K_{S}$ colour magnitude diagrams of our target clusters. All stars within the box are used to determine the mean RC magnitude in the clusters.}
    \label{fig:CMD}
\end{figure*}

We used Palma's catalogue value as cluster radii and chose stars in circular regions as cluster stars. For stars within the circles, we plotted colour-magnitude diagrams and decided to use stars with $17.5 < K_{S} < 15.0$ and $0.2 < J - K_{S} < 0.8$ to fit the luminosity function of the RC stars (Figure~\ref{fig:CMD}). We fitted the magnitude distribution of the stars with a function of the form
\begin{equation}
N(\lambda) = a + bm_{\lambda}+cm_{\lambda}^2 + d \exp\left[-\frac{(m_{\lambda}^{RC} - m_{\lambda})^2}{2\sigma_{\lambda}^2}\right],
\label{eq:fitting}
\end{equation}
where $\lambda$ is a passband ($JHK_{S}$). The first three terms represent the background distribution of red giant branch stars. The Gaussian term represents the RC stars distribution, where $m_{\lambda}^{\mathrm{RC}}$ is the mean magnitude that we desire to acquire and $\sigma_{\lambda}$ is the standard deviation of the RC stars. The uncertainty in $m_{\lambda}^{RC}$ is standard error and calculated by
\begin{equation}
\mathrm{standard\ error} = \frac{\sigma_{\lambda}}{\sqrt{N_{\mathrm{RC}}}},
\end{equation}
where $N_{\mathrm{RC}}$ is the number of RC stars calculated by
\begin{equation}
N_{\mathrm{RC}} = \sqrt{2\pi} \sigma_{\lambda} \times 10d.
\end{equation}
Factor 10 corresponds to the width of histogram bin 0.1~mag. A typical standard error of the RC mean magnitudes for the clusters is 0.015~mag. Then, reddening was corrected using $E(B - V)$ values from Palma's catalogue and the \citet{CCM1989} extinction law. We only used clusters that had a significant excess of red clump stars. As a consequence, 10~clusters were used to investigate the magnitude of RC stars. The parameters of the clusters used in this work are listed in Table~\ref{tab:cluster}.

\begin{table*}
\caption{Cluster Information}
\label{tab:cluster}
\scalebox{0.9}{
\begin{tabular}{lcccccccc}
\hline
Cluster name & Radius (arcmin) & $E(B - V)$ & Age (Gyr) & [Fe/H] & $m_{J}^{a}$ & $m_{H}^{a}$ & $m_{K_s}^{a}$ & $m_{J} - m_{K_s}^{a}$\\
\hline
KMHK 21 & 0.75 & 0.040 & 1.60 $\pm$ 0.30 & -0.50 $\pm$ 0.30 & 17.485 $\pm$ 0.017 & 16.962 $\pm$ 0.029 & 16.860 $\pm$ 0.027 & 0.625 $\pm$ 0.032\\
KMHK 337 & 0.68 & 0.020 & 2.00 $\pm$ 0.20 & -0.65 $\pm$0.20 & 17.381 $\pm$ 0.015 & 16.900 $\pm$ 0.017 & 16.833 $\pm$ 0.013 & 0.548 $\pm$ 0.020\\
ESO 85-72 & 0.85 & 0.030 & 2.20 $\pm$ 0.30 & -0.65 $\pm$ 0.30 & 17.279 $\pm$ 0.025 & 16.806 $\pm$ 0.024 & 16.799 $\pm$ 0.017 & 0.480 $\pm$ 0.030\\ 
NGC 1997 & 0.90 & 0.040 & 2.60 $\pm$ 0.50 & -0.70 $\pm$ 0.20 & 17.374 $\pm$ 0.014 & 16.889 $\pm$ 0.015 & 16.857 $\pm$ 0.015 & 0.517 $\pm$ 0.020\\
IC 2140 & 1.15 & 0.111 & $2.50^{+0.60}_{-0.50}$ & $-0.84^{+0.22}_{-0.18}$ & 17.337 $\pm$ 0.020 & 16.860 $\pm$ 0.019 & 16.794$\pm$ 0.019 & 0.543 $\pm$ 0.027\\
KMHK 1281 & 0.80 & 0.050 & 2.00 $\pm$ 0.40 & -0.90 $\pm$ 0.20 & 17.272 $\pm$ 0.011 & 16.802 $\pm$ 0.011 & 16.750 $\pm$ 0.014 & 0.522 $\pm$ 0.018\\
NGC 2161 & 1.15 & 0.130 & 1.10 $\pm$ 0.30 & -0.70$^b$ & 17.374 $\pm$ 0.012 & 16.961 $\pm$ 0.013 & 16.896 $\pm$ 0.012 & 0.478 $\pm$ 0..017\\
NGC 2155 & 1.20 & 0.050 & 3.20 $\pm$ 0.60 & -0.90 $\pm$ 0.20 & 17.254 $\pm$ 0.011 & 16.771 $\pm$ 0.010 & 16.728 $\pm$ 0.018 & 0.527 $\pm$ 0.021\\
ESO 121-3 & 1.05 & 0.030 & 8.50 $\pm$ 0.30 & -1.05 $\pm$ 0.30 & 17.308 $\pm$ 0.018 & 16.900 $\pm$ 0.015 & 16.827 $\pm$ 0.024 & 0.480 $\pm$ 0.030\\
NGC 2213 & 1.05 & 0.116 & 1.50$^b$ &  -0.40 $\pm$ 0.15 & 17.427 $\pm$ 0.012 & 16.906 $\pm$ 0.016 & 16.915  $\pm$ 0.010 & 0.512 $\pm$ 0.016\\
\hline
\multicolumn{8}{l}{$^a$ Interstellar extinction was corrected by \citet{CCM1989} extinction law.}\\
\multicolumn{8}{l}{$^b$ The uncertainties are not described in \citet{PGC2016}}.\\
\end{tabular}
}
\end{table*}

\section{Results and discussion}\label{sec:Results}
The $JHK_{S}$ magnitude distribution of the RC stars and fitting results are shown in Figures~\ref{fig:RC_fitting} and \ref{fig:RC_fitting2}. RC magnitudes derived from the fittings using equation~(\ref{eq:fitting}) are listed in Table \ref{tab:cluster}. It is possible that the mean magnitudes obtained with the fittings are sensitive to the bin size of $m_{\lambda}$. We changed the bin size to narrower value (0.05) and wider value (0.20) to check the effect of the bin size. The results are shown in Table~\ref{tab:Mag_diff}. As can be seen in Table~\ref{tab:Mag_diff}, most of the obtained magnitudes for the narrower or wider bin size are consistent with those for the bin size of 0.10. The mean differences of magnitudes between the bin sizes of 0.05 and 0.10, and 0.10 and 0.20 are small, 0.013~mag and 0.024~mag, respectively. For the wider bin size, some clusters show the relatively large difference compared to the standard errors. However, this difference does not much affect the obtained trends and our discussion.

\begin{figure*}
    \includegraphics{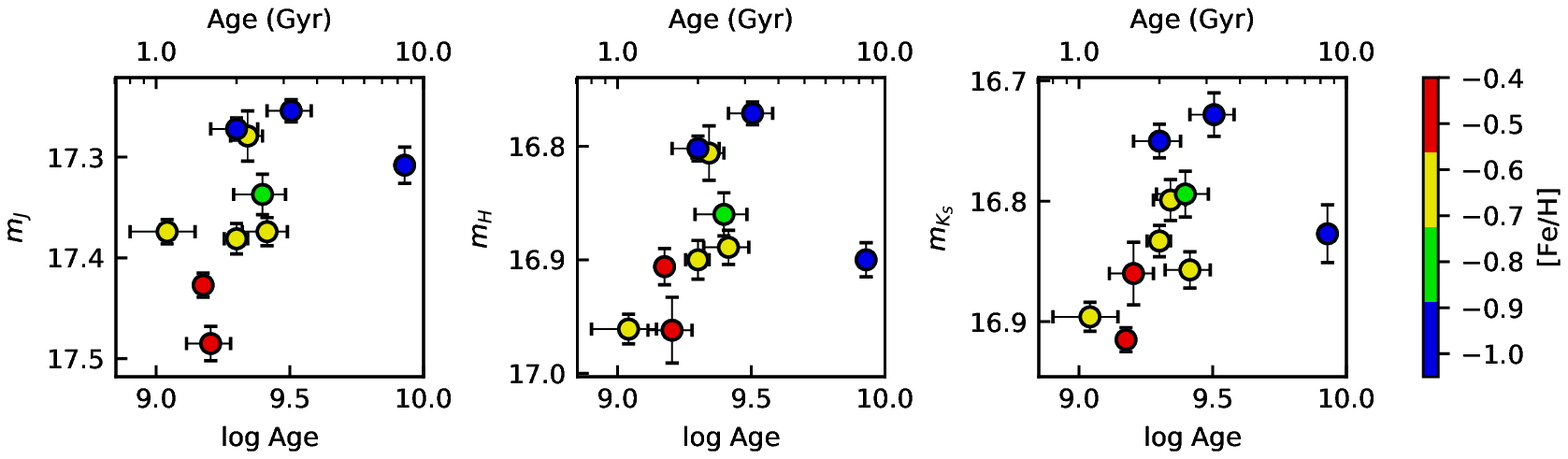}
    \caption{Mean RC magnitude versus age in the $J$- (left), $H$- (centre), and $K_{S}$-bands (right).  Metallicity difference is shown by colour scale.}
    \label{fig:mag_age}
\end{figure*}

\begin{figure*}
    \includegraphics{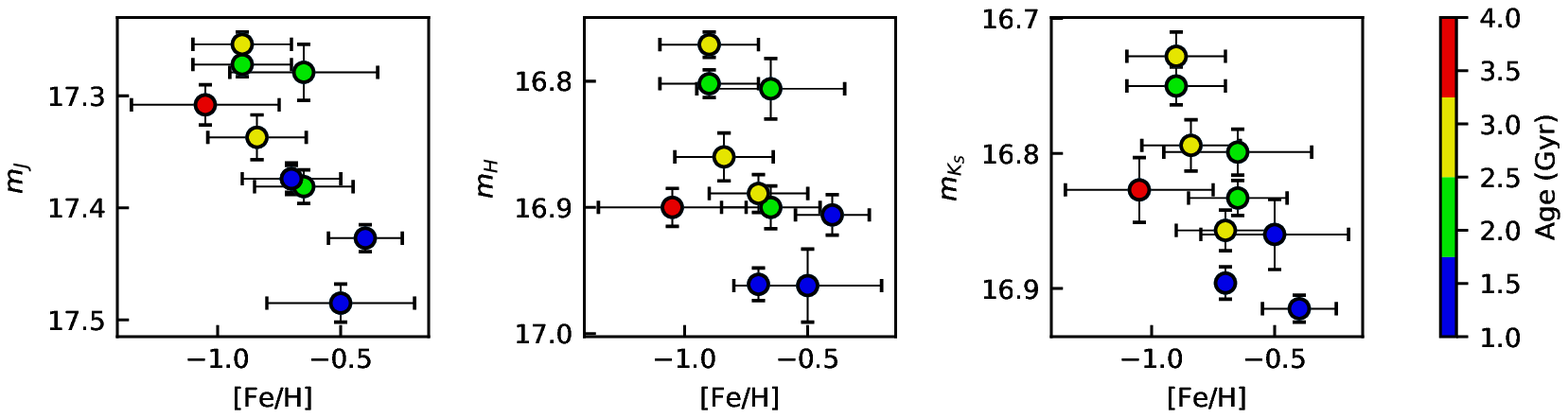}
    \caption{Mean RC magnitude versus metallicity in the $J$- (left), $H$- (centre), and $K_{S}$-bands (right). Age difference is shown by colour scale.}
    \label{fig:mag_metallicity}
\end{figure*}

\begin{figure*}
    \includegraphics{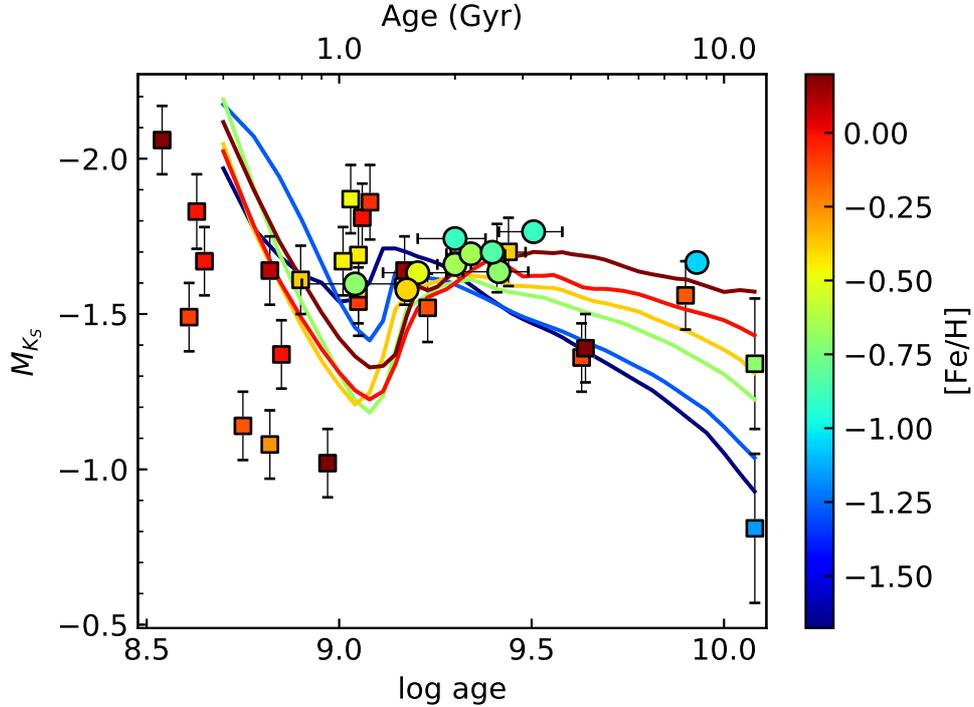}
    \caption{The comparison of $K_{S}$-band absolute magnitudes between model predictions and observations. Solid lines represent model prediction for some different metallicities from table~1 of \citet{SG2002}. Circles represent the IRSF data, and squares are data from \citet{vHG2007}. Metallicity difference is illustrated by colour scale.}
    \label{fig:Model_comparison}
\end{figure*}

\begin{figure*}
    \includegraphics{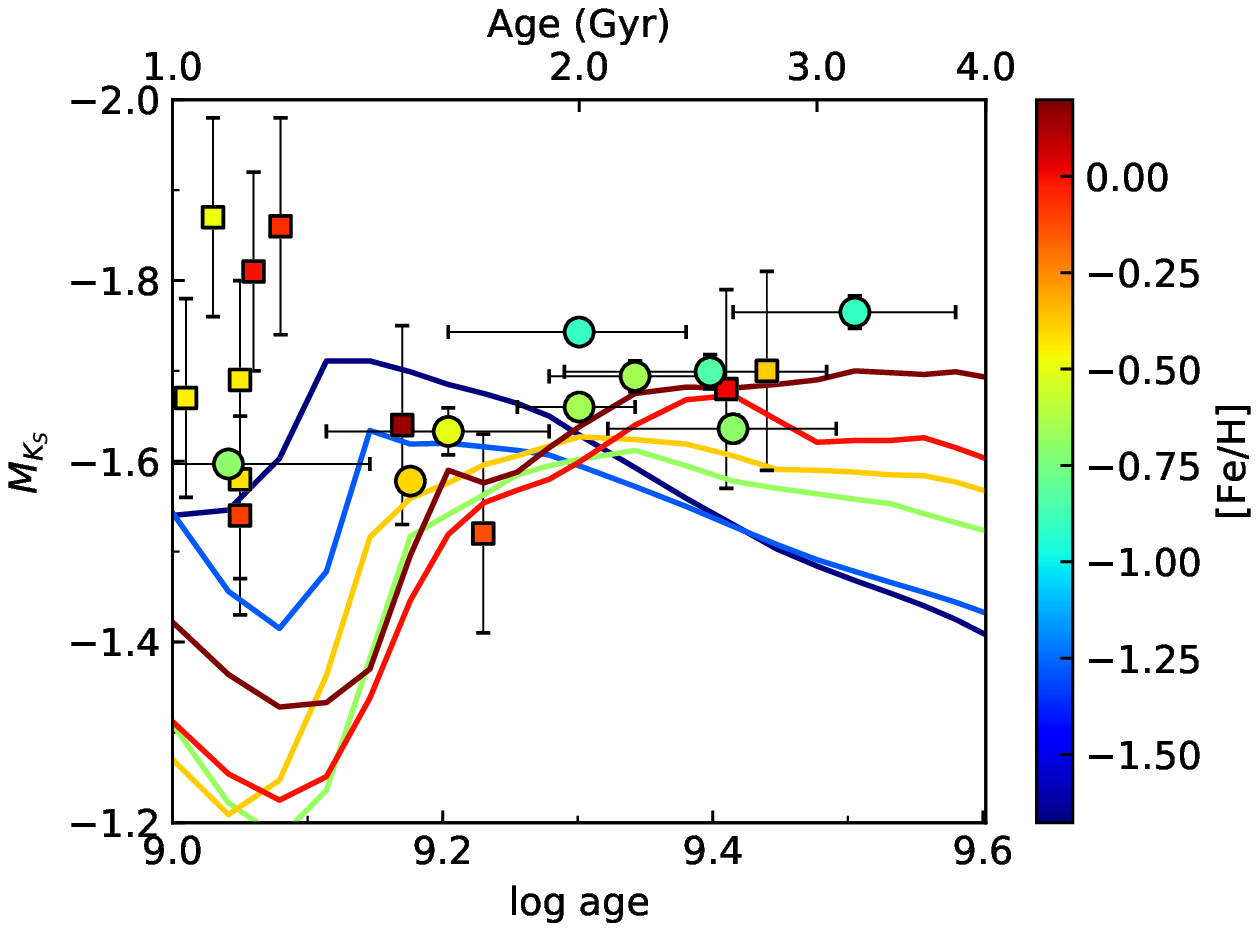}
    \caption{The same plot as Figure \ref{fig:Model_comparison} but changing the plot range to focus on the samples between 1 and 4~Gyr.}
    \label{fig:Model_comparison2}
\end{figure*}

The $m_{\lambda}$ values for the star clusters in our sample versus age are plotted in Figure~\ref{fig:mag_age}, and the $m_{\lambda}$ versus metallicity are plotted in Figure~\ref{fig:mag_metallicity}. Figures~\ref{fig:Model_comparison} and  \ref{fig:Model_comparison2} present comparison of our $K_{S}$-band results with data from \citet{vHG2007} and the model of \citet{SG2002}. When we convert apparent magnitudes of our results to absolute magnitudes, we use the distance modulus to the LMC \citep[18.493~mag,][]{PGG2013}. We assume that all clusters in the LMC are at the same distance because we cannot obtain the distance information on each cluster from Palma's catalogue.

\begin{figure*}
    \includegraphics{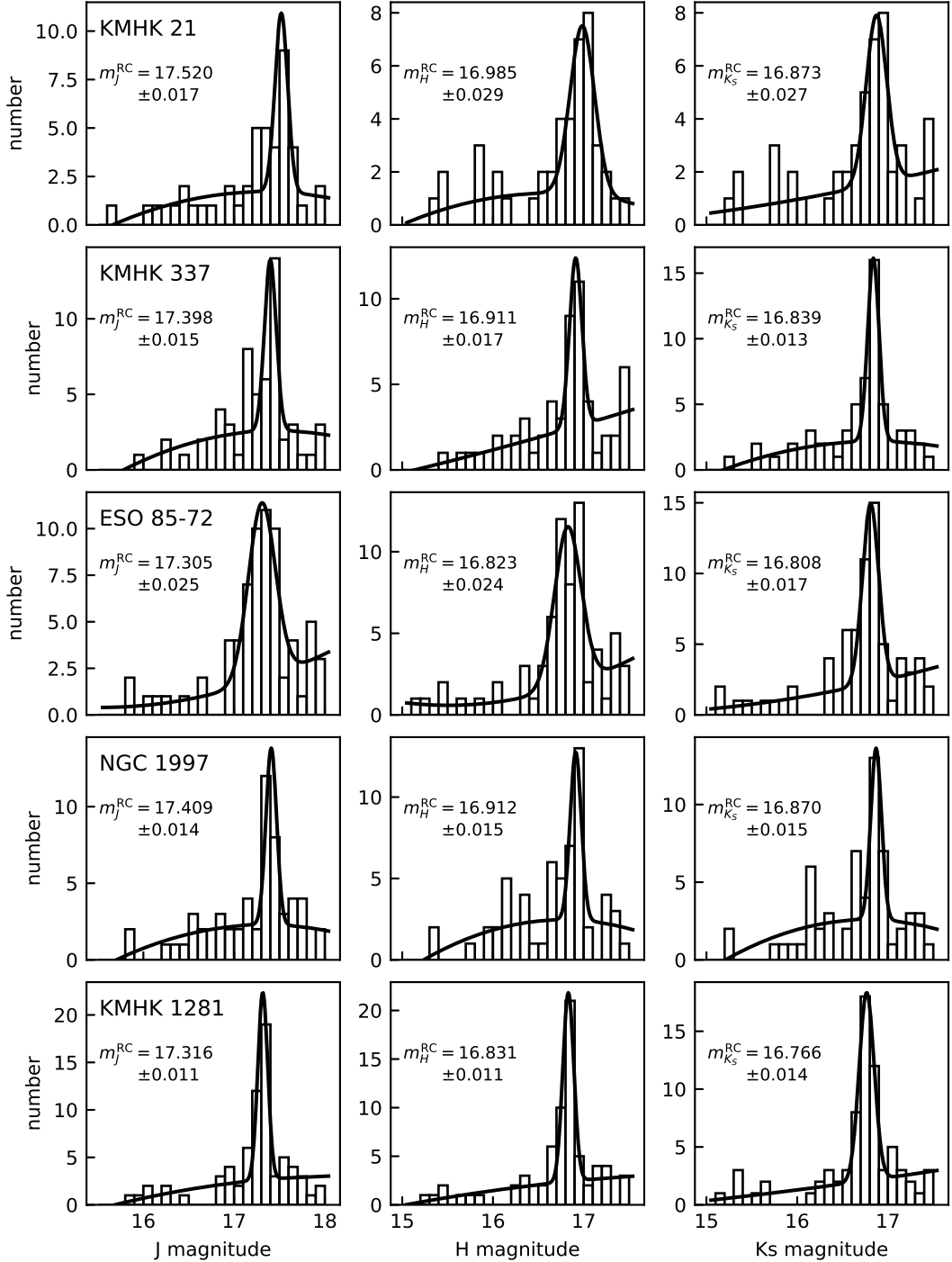}
    \caption{The distribution of the RC stars as a function of their magnitudes in the $J$- (left), $H-$ (centre), and $K_{S}$-bands (right). The fitting results with the equation~(\ref{eq:fitting}) are shown by black lines.}
    \label{fig:RC_fitting}
\end{figure*}

\begin{figure*}
    \includegraphics{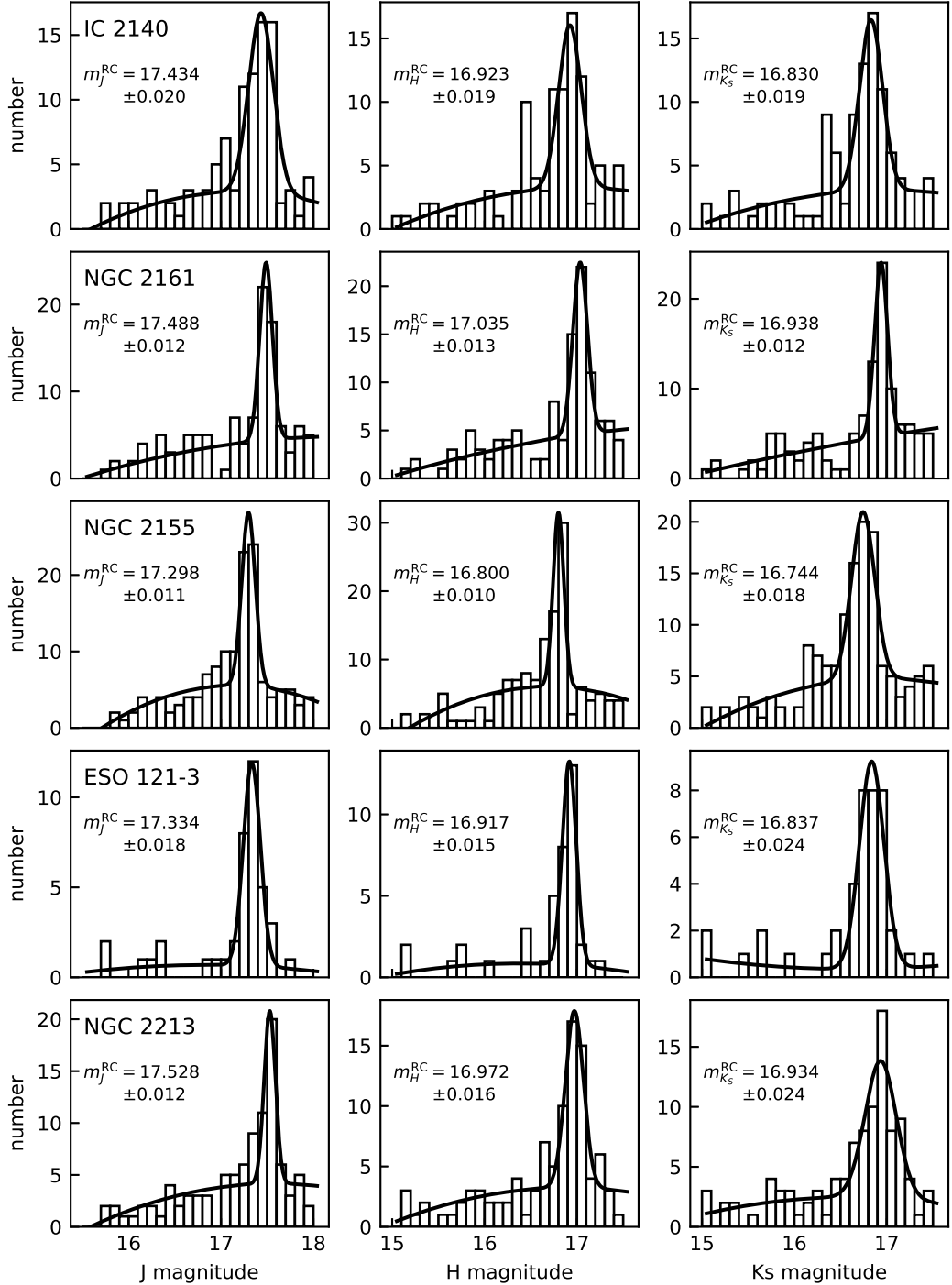}
    \caption{Continued from Figure \ref{fig:RC_fitting}.}
    \label{fig:RC_fitting2}
\end{figure*}

\begin{table*}
\caption{The difference of magnitudes for three bin sizes 0.05, 0.10, and 0.20}
\label{tab:Mag_diff}
\scalebox{0.9}{
\begin{tabular}{lcccccc}
\hline
Cluster name & Filter & Bin size of 0.10 & Bin size of 0.05 & Difference between 0.10 and 0.05 & Bin size of 0.20 & Difference between 0.10 and 0.20\\
\hline
 & $J$ & 17.485 $\pm$ 0.017 & 17.457 $\pm$ 0.029 & 0.028 & 17.407 $\pm$ 0.032 & 0.078\\
KMHK 21 & $H$ & 16.962 $\pm$ 0.029 & 16.947 $\pm$ 0.033 & 0.015 & 16.946 $\pm$ 0.028 & 0.016\\
& $K_{S}$ & 16.860 $\pm$ 0.027 & 16.888 $\pm$ 0.015 & 0.028 & 16.846 $\pm$ 0.034 & 0.014\\
& $J$ & 17.381 $\pm$ 0.015 & 17.422 $\pm$ 0.011 & 0.041 & 17.351 $\pm$ 0.044 & 0.030\\
KMHK 337 & $H$ & 16.900 $\pm$ 0.017 & 16.897 $\pm$ 0.018 & 0.003 & 16.873 $\pm$ 0.027 & 0.027\\
& $K_{S}$ & 16.833 $\pm$ 0.013 & 16.820 $\pm$ 0.009 & 0.013 & 16.757 $\pm$ 0.019 & 0.076\\
& $J$ & 17.279 $\pm$ 0.025 & 17.308 $\pm$ 0.021 & 0.029 & 17.266 $\pm$ 0.027 & 0.013\\
ESO 85-72 & $H$ & 16.806 $\pm$ 0.024 & 16.828 $\pm$ 0.021 & 0.022 & 16.812 $\pm$ 0.028 & 0.006\\
& $K_{S}$ & 16.799 $\pm$ 0.017 & 16.799 $\pm$ 0.020 & 0.000 & 16.740 $\pm$ 0.018 & 0.059\\
& $J$ & 17.374 $\pm$ 0.014 & 17.346 $\pm$ 0.017 & 0.028 & 17.374 $\pm$ 0.041 & 0.000\\
NGC 1997 & $H$ & 16.889 $\pm$ 0.015 & 16.888 $\pm$ 0.007 & 0.001 & 16.850 $\pm$ 0.021 & 0.039\\
& $K_{S}$ & 16.857 $\pm$ 0.015 & 16.873 $\pm$ 0.008 & 0.016 & 16.774 $\pm$ 0.029 & 0.083\\
& $J$ & 17.337 $\pm$ 0.020 & 17.354 $\pm$ 0.017 & 0.017 & 17.318 $\pm$ 0.013 & 0.019\\
IC 2140 & $H$ & 16.860 $\pm$ 0.019 & 16.869 $\pm$ 0.020 & 0.009 & 16.857 $\pm$ 0.017 & 0.003\\
& $K_{S}$ & 16.794$\pm$ 0.019 & 16.802 $\pm$ 0.019 & 0.008 & 16.797 $\pm$ 0.021 & 0.003\\
& $J$ & 17.272 $\pm$ 0.011 & 17.276 $\pm$ 0.010 & 0.004 & 17.252 $\pm$ 0.017 & 0.020\\
KMHK 1281 & $H$ & 16.802 $\pm$ 0.011 & 16.805 $\pm$ 0.011 & 0.003 & 16.758 $\pm$ 0.017 & 0.044\\
& $K_{S}$ & 16.750 $\pm$ 0.014 & 16.760 $\pm$ 0.012 & 0.010 & 16.763 $\pm$ 0.018 & 0.013\\
& $J$ & 17.374 $\pm$ 0.012 & 17.377 $\pm$ 0.011 & 0.003 & 17.397 $\pm$ 0.015 & 0.023\\
NGC 2161 & $H$ & 16.961 $\pm$ 0.013 & 16.949 $\pm$ 0.011 & 0.012 & 16.946 $\pm$ 0.018 & 0.015\\
& $K_{S}$ & 16.896 $\pm$ 0.012 & 16.900 $\pm$ 0.011 & 0.004 & 16.911 $\pm$ 0.018 & 0.015\\
& $J$ & 17.254 $\pm$ 0.011 & 17.249 $\pm$ 0.012 & 0.005 & 17.255 $\pm$ 0.017 & 0.001\\
NGC 2155 & $H$ & 16.771 $\pm$ 0.010 & 16.801 $\pm$ 0.008 & 0.030 & 16.764 $\pm$ 0.014 & 0.007\\
& $K_{S}$ & 16.728 $\pm$ 0.018 & 16.745 $\pm$ 0.014 & 0.017 & 16.712 $\pm$ 0.015 & 0.016\\
& $J$ & 17.308 $\pm$ 0.018 & 17.305 $\pm$ 0.015 & 0.003 & 17.313 $\pm$ 0.021 & 0.005\\
ESO 121-3 & $H$ & 16.900 $\pm$ 0.015 & 16.892 $\pm$ 0.022 & 0.008 & 16.896 $\pm$ 0.023 & 0.004\\
& $K_{S}$ & 16.827 $\pm$ 0.024 & 16.840 $\pm$ 0.022 & 0.013 & 16.843 $\pm$ 0.026 & 0.016\\
& $J$ & 17.427 $\pm$ 0.012 & 17.423 $\pm$ 0.009 & 0.004 & 17.357 $\pm$ 0.018 & 0.070\\
NGC 2213 & $H$ & 16.906 $\pm$ 0.016 & 16.917 $\pm$ 0.016 & 0.011 & 16.909 $\pm$ 0.019 & 0.003\\
& $K_{S}$ & 16.915  $\pm$ 0.010 & 16.899 $\pm$ 0.013 & 0.016 & 16.930 $\pm$ 0.027 & 0.015\\
\hline
\end{tabular}
}
\end{table*}

\subsection{Age dependence}
Apparent magnitudes $m_{J}$, $m_{H}$, $m_{K_{S}}$ show the same trends that older RC stars are brighter within 1-3~Gyr, and much older RCs are slightly fainter (Figure~\ref{fig:mag_age}). This trend still can be seen for RC stars with similar metallicity. These are the same trends seen in $VJK$-band data \citep[although they have only three or four data points in the same age range]{PS2003} and in $K$-band data \citep{GS2002,vHG2007} for the Milky Way clusters. Theoretical models \citep{SG2002,G2016} predict very similar trends to our results, although the individual values of absolute magnitudes are slightly different (Figure~\ref{fig:Model_comparison}). For RC stars between 1.5 and 2.8~Gyr, both our observational data and data from \citet{vHG2007} are consistent with the theoretical models within 0.1~mag. NGC~2155 (3.2~Gyr) shows slightly large deviation between observational data and the model prediction by \citet{SG2002}, but in good agreement with the model by \citet{G2016}.

The younger cluster NGC~2161 (1.1~Gyr) shows brighter magnitude compared to the theoretical models. For RC stars younger than 1.5~Gyr, strong age dependence of absolute magnitudes is predicted. Even small age difference leads to the large difference of absolute magnitudes for these young RC stars. We note that the observational data from \citet{vHG2007} also shows systematically brighter results around 1~Gyr. The old cluster ESO~121-3 (8.5~Gyr) also shows a brighter value than the model predictions. One possible reason is that the number of fainter RC stars is underestimated because the photometric data is not deep enough. The number of stars contained in ESO~121-3 is smaller than those contained in other clusters (Figure~\ref{fig:CMD}). However, the difference between the observational data and the theoretical models is about 0.3--0.4~mag. The widths of RC magnitudes in other star clusters or the $JH$-band magnitudes in ESO~121-3 are not wider than 0.3~mag. Even if we only detect the bright tail of RC distribution, the peak of magnitude distribution can only change 0.2~mag at most.  Moreover, we added artificial stars to the reduced image using the \textsc{iraf/daophot/addstar} task, and confirmed that the completeness is nearly 100~per~cent at around 17.0~mag in $K_{S}$-band. Thus, the underestimation of faint RC stars would not change the result so much. Other reasons may be needed to explain the deviation.

\begin{figure*}
    \includegraphics{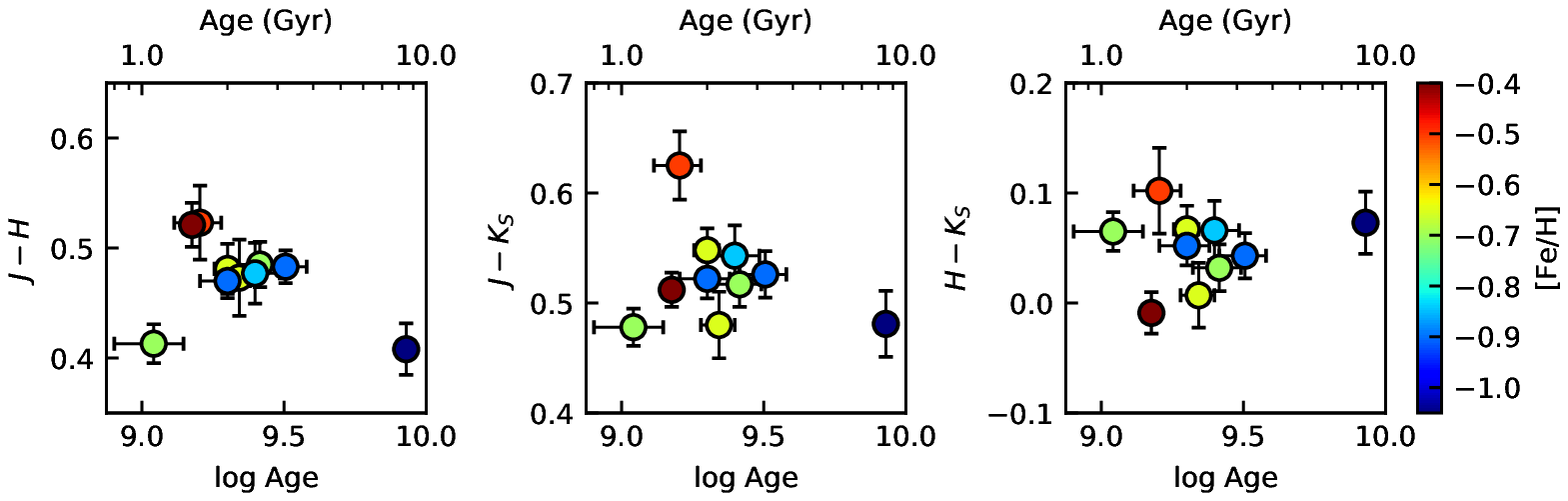}
    \caption{(Left) $J - H$, (centre) $J - K_{S}$, and (right) $H - K_{S}$ colours of RC stars as a function of age. Metallicity difference is shown by colour scales.}
    \label{fig:RC_color}
\end{figure*}

The age dependence of $J - H$, $J - K_{S}$, and $H - K_{S}$ colours is shown in Figure \ref{fig:RC_color}. $J - H$, and $J - K_{S}$ colours show weak age dependence that the older RC stars have redder colours between 1 and 3~Gyr. On the other hand, no significant age dependence can be seen in $H - K_{S}$ colour. These trends are similar to the model predictions. The theoretical models predict that younger RC stars have bluer colours and older RC stars have redder colours. In addition, age dependence becomes weaker in longer wavelengths. Comparing $V - K$ ($\sim$0.5~mag difference for similar ages and metallicities) and $I - K$ ($\sim$0.3~mag) colours of the model prediction by \citet{SG2002}, our observational data show weak age dependence ($\sim$0.1~mag).

Most of our sample clusters have the age of 1--3~Gyr where theoretical models predict strong age dependence. This age dependence is clearly confirmed in our study, thanks to many samples in this narrow age range.  For very young (< 1 Gyr) RC stars, theoretical models predict that younger RC stars have brighter magnitudes. However, such young star clusters do not meet our target selection criteria; they are expected to be very small or exist in the bar region, so we cannot investigate the trend in this study. The VMC survey will play an important role to investigate the dependence of very young RC stars.

\subsection{Metallicity dependence}
We can see metallicity dependence on $m_{J}$, $m_{H}$ and $m_{K_{S}}$ (Figure~\ref{fig:mag_metallicity}). The predicted metallicity dependence in the $K$-band is only 0.1--0.2~mag around 2~Gyr, and this is smaller than those in the shorter wavelengths. The expected trend is that RC stars with lower metallicity have brighter magnitudes; this trend can be seen in our results (Figure~\ref{fig:mag_metallicity}). The magnitude difference found in our sample is about 0.2~mag, and this matches very well with the theoretical prediction. 

\begin{figure*}
    \includegraphics{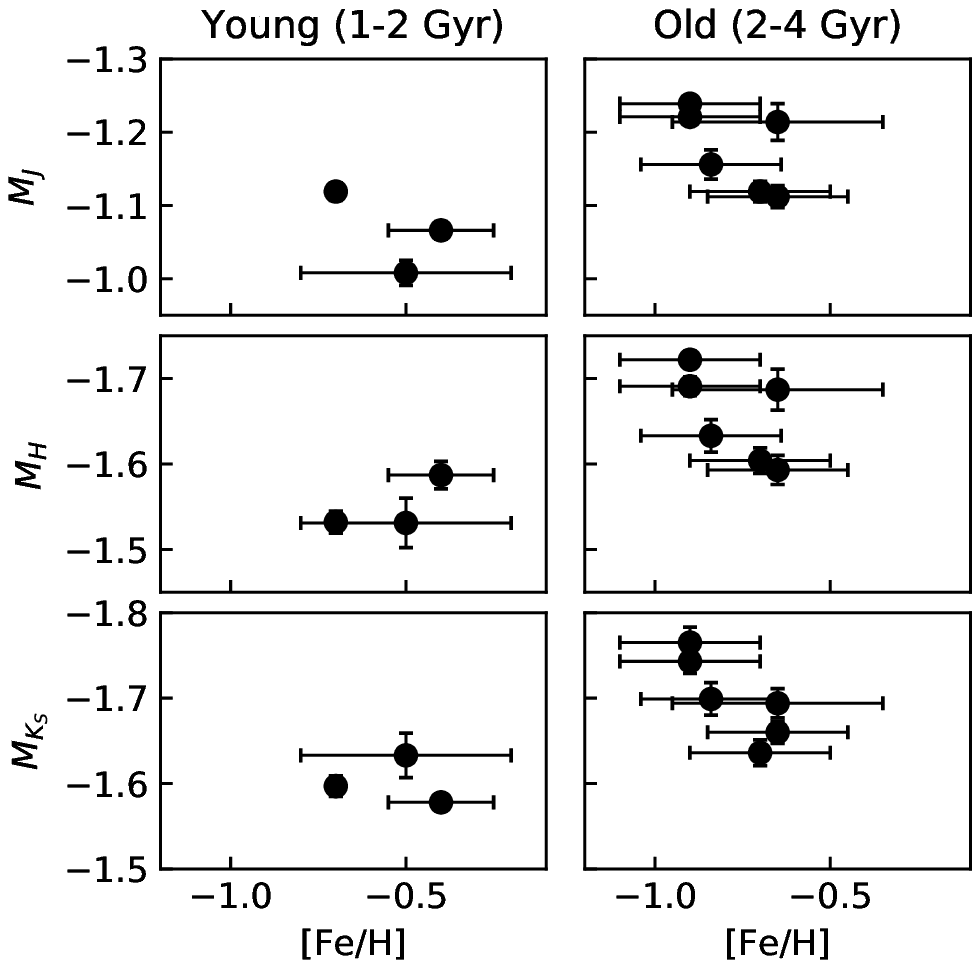}
    \caption{Mean RC magnitude versus metallicity for (left) younger (1~Gyr $\leq$ Age $<$ 2~Gyr) and (right) older (2~Gyr $\leq$ Age $<$ 4~Gyr) RC stars in the $J$- (uppar), $H$- (center), and $K_{S}$-bands (bottom).}
    \label{fig:young_and_old}
\end{figure*}

The predicted metallicity dependence is different in two models between 2 and 4~Gyr. \citet{SG2002} predicted that metal-rich RC stars have brighter $K_{S}$-band absolute magnitudes. On the other hand, \citet{G2016} presented a contrary result that metal-poor RC stars have fainter $K_{S}$-band absolute magnitudes within the age range. To compare these model predictions with our observational data, we divided our samples into younger RC stars (1--2~Gyr) and older RC stars (2--4~Gyr). Figure~\ref{fig:young_and_old} provides metallicity dependence of RC absolute magnitudes for the divided samples. In this plot, the $K_{S}$-band absolute magnitudes show the trend that metal-rich RC stars have fainter magnitudes between 2 and 4~Gyr. This result is consistent with the prediction by \citet{G2016}, although our sample shows slightly stronger metallicity dependence. The $J$- and $H$-band data also show metal-rich RC stars have fainter absolute magnitudes.

\begin{figure*}
    \includegraphics{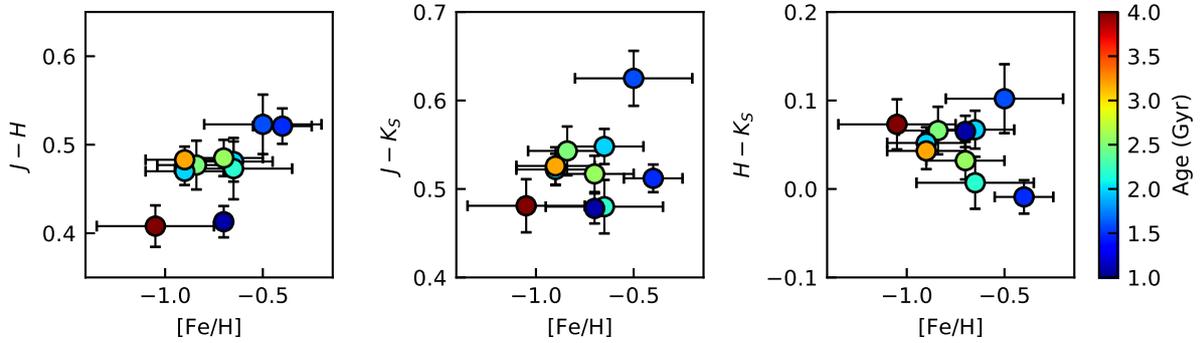}
    \caption{(Left) $J - H$, (centre) $J - K_{S}$, and (right) $H - K_{S}$ colours of RC stars as a function of metallicity. Age difference is shown by colour scales.}
    \label{fig:RC_color_metal}
\end{figure*}

The metallicity dependence of $J - H$, $J - K_{S}$, and $H - K_{S}$ colours is presented in Figure~\ref{fig:RC_color_metal}. $J - H$ colour shows the trend that metal-rich RC stars have redder colours. This trend is consistent with the model predictions for $V - K$ and $I - K$ colours. The difference of $J - H$ colours is $\sim$0.1~mag and smaller than that predicted for $V - K$ ($\sim$0.5~mag) and $I - K$ ($\sim$0.3~mag) with similar metallicities. This result matches the model predictions and previous observational studies that metallicity dependence becomes weaker at longer wavelengths. We can see no strong metallicity dependence of $J - K_{S}$ and $H - K_{S}$ colours.

\subsection{Absolute magnitude}
The averages of the apparent RC magnitudes for our 10 clusters are $17.349 \pm 0.023$~mag, $16.876 \pm 0.021$~mag, and $16.826  \pm 0.019$~mag for $m_{J}$, $m_{H}$, and $m_{K_{S}}$, respectively. Considering the distance modulus to the LMC \citep[$18.493 \pm 0.008 \pm 0.047$~mag,][]{PGG2013}, absolute magnitudes of RC stars, $M_{J}$, $M_{H}$, and $M_{K_{S}}$ become $-1.144 \pm 0.023$ (this work's error) $\pm 0.008$ (Pietrzy{\'n}ski's statistical) $\pm 0.047$ (Pietrzy{\'n}ski's systematic)~mag, $-1.617 \pm 0.021 \pm 0.008 \pm 0.047$~mag, and $-1.667 \pm 0.019 \pm 0.008 \pm 0.047$~mag, respectively. These results give good agreement with previous work in the $K_{S}$-band, but 0.1-0.2~mag brighter than previous studies derived from RC stars in the solar neighbourhood or Kepler field for the $J$- and $H$-bands \citep{LJP2012, CCZ2017, HLB2017, RBA2018}. \citet{CGB2003} derived a slightly different distance modulus to the LMC of $18.42 \pm 0.07$ using RC stars. If we use the value, absolute magnitudes of RC stars $M_{J}$, $M_{H}$, and $M_{K_{S}}$ change to $-1.071 \pm 0.023 \pm 0.07$, $-1.544  \pm 0.021 \pm 0.07$, and $-1.594 \pm 0.019 \pm 0.07$. These values have better agreement in the $J$- and $H$-bands.

\citet{LJP2012} pointed that the distance modulus to the LMC derived from $J$-band RC magnitudes was about 0.1 mag smaller than that from $H$- or $K_{S}$-band RC magnitudes. They obtained $JHK_{S}$ absolute magnitudes of RC stars by combining their photometric results with \textit{Hipparcos} parallaxes. They compared these absolute magnitudes with $JHK$ magnitudes of LMC RC stars, and found that the distance modulus derived from $J$-band data is relatively small. Our data show equivalent results that mean $J$-band absolute magnitude is brighter than that for RC stars in the solar neighborhood when we use the same distance modulus. \citet{LJP2012} suggested that the discrepancy was probably caused by the population effect. Both theoretical models and observations have confirmed that the population effects become stronger in the shorter wavelengths. $J - K_{S}$ colours of RC stars in our target clusters are 0.47-0.63 mag (Table \ref{tab:cluster}), and bluer than those in solar neighborhood \citep[0.629;][]{LJP2012} or in Baade's Window \citep[0.68;][]{GRZ2012}. Theoretical models predict that metal-poor RC stars have bluer colour than metal-rich RC stars because of the stronger population effect in the shorter wavelengths. Metallicities of RC stars in our target clusters are lower than those in solar neighborhood or in Baade's Window. This results support that $J$-band population effect is stronger than $K_{S}$-band. Figures~\ref{fig:mag_age} and \ref{fig:mag_metallicity} also indicate that the population effect in the $J$-band is slightly stronger than that in the $K_{S}$-band. In $J$-band, the difference between the brightest RC stars (NGC~2155) and the faintest ones (KMHK~21) is 0.231~mag. In $K_{S}$-band, the difference between the brightest (NGC~2155) and faintest (NGC~2213) RC stars becomes smaller (0.187~mag). $J - K_{S}$ colour also shows age dependence (Figure~\ref{fig:RC_color}). Therefore, the discrepancy in the $J$-band is probably caused by the population effects.

\subsection{RC stars as a standard candle}
So far, the population effect of RC absolute magnitudes has been corrected using theoretical models by \citet{GS2001} and \citet{SG2002}. To obtain the correction of the population effect from observational data, we fitted our data using least squares method and following function
\begin{equation}
M_{\lambda} = a \log t  + b [\mathrm{Fe}/\mathrm{H}] + c,
\label{eq:Mag_fit}
\end{equation}
where $t$ is the age (yr) of the star clusters. We excluded ESO~121-3 from the fitting because the age of this cluster is very old and the theoretically expected behaviour of RC magnitudes is completely different from young ones. As the best fit result, we obtained
\begin{align}
M_{J} = (-0.098 \pm 0.153) \log t + (0.315 \pm 0.126) [\mathrm{Fe}/\mathrm{H}] \notag \label{eq:M_J}\\
- (0.006 \pm 1.378),\\
M_{H} = (-0.277 \pm 0.143) \log t + (0.136 \pm 0.117) [\mathrm{Fe}/\mathrm{H}] \notag \label{eq:M_H}\\
+ (1.048 \pm 1.284),\\
M_{K_{S}} = (-0.185 \pm 0.100) \log t + (0.220 \pm 0.082) [\mathrm{Fe}/\mathrm{H}]\notag \label{eq:M_K}\\
+ (0.206 \pm 0.898).
\end{align}
The adjusted coefficients of determination (adjusted $R^{2}$) are 0.568, 0.537, and 0.733, and these values are calculated by
\begin{equation}
\mathrm{adjusted}\ R^{2} = 1 - \frac{\sum_{i}(M_{\lambda, i} - f_{i})^2 / (N - p - 1)}{\sum_{i}(M_{\lambda, i} - \overline{M_{\lambda}})^2 / (N  - 1)},
\end{equation}
where $M_{\lambda}$ is the absolute magnitudes of observational data in the $\lambda$-band, $f$ is the absolute magnitudes derived from equations~(\ref{eq:M_J})--(\ref{eq:M_K}), $\overline{M_{\lambda}}$ is the mean value of $M_{\lambda}$, $N$ is the number of sample clusters (nine in this time), and $p$ is the number of explanatory variables (three in this time). The adjusted $R^{2}$ gets closer to one for the better fitting. If the adjusted $R^{2}$ is negative, the fitting is worse than just taking the average value. We also fitted our data with higher order polynomial functions but adjusted $R^{2}$ values are worse than fitting with equation~(\ref{eq:Mag_fit}). To evaluate the order of model equation, we performed leave-one-out cross-validation for equation~(\ref{eq:Mag_fit}) and higher order polynomial functions
\begin{align}
M_{\lambda} &= a \log t + b(\log t)^2  + c [\mathrm{Fe}/\mathrm{H}] + d, \label{eq:Mag_fit_2}\\
M_{\lambda} &= a \log t  + b [\mathrm{Fe}/\mathrm{H}] + c[\mathrm{Fe}/\mathrm{H}]^2 + d, \label{eq:Mag_fit_3}\\
M_{\lambda} &= a \log t  + b(\log t)^2 + c [\mathrm{Fe}/\mathrm{H}] + d [\mathrm{Fe}/\mathrm{H}]^2 + e.
\label{eq:Mag_fit_4}
\end{align}
Eight clusters are used as a training set and the other cluster is used as a test set. We calculated the differences of absolute magnitudes between observational data and calculated values from equations (\ref{eq:Mag_fit}), (\ref{eq:Mag_fit_2}) - (\ref{eq:Mag_fit_4}) for the test sets, and derived root mean squares (RMSs). The obtained RMSs are shown in Table \ref{tab:Mag_RMS}. As can be seen in Table \ref{tab:Mag_RMS}, the RMSs are similar or become larger for higher order polynomial functions. In such a situation, it is preferred to select the simplest model whose error is within one standard error of the minimal error (one standard error rule). Therefore, we chose the form of equation~(\ref{eq:Mag_fit}) as the best model.

\begin{figure*}
    \includegraphics{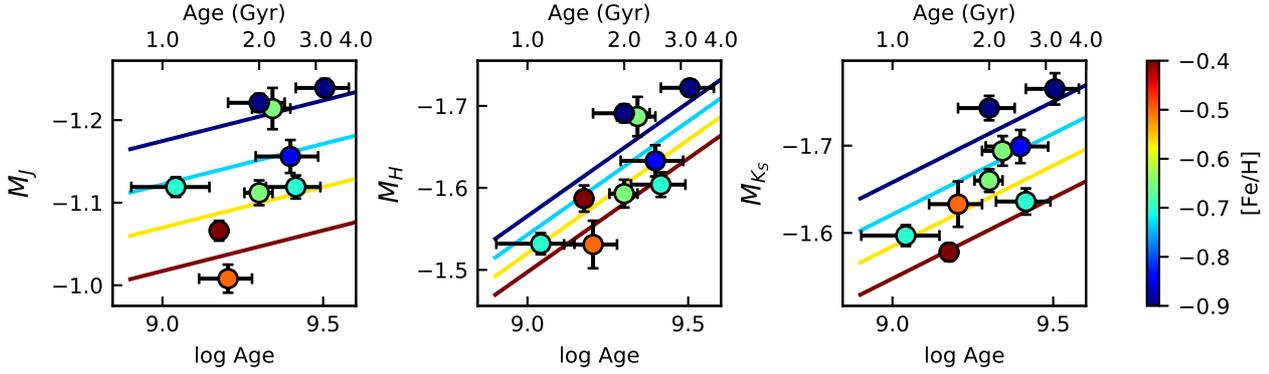}
    \caption{Mean RC magnitude versus age in the $J$- (left), $H$- (centre), and $K_{S}$-band (right). The best-fit relations are also plotted for four metallicities (from blue to brown, $-$0.90, $-$0.73, $-$0.57, $-$0.40~dex). Metallicity difference is shown by colour scales.}
    \label{fig:RC_age_mag_fit}
\end{figure*}

\begin{figure*}
    \includegraphics{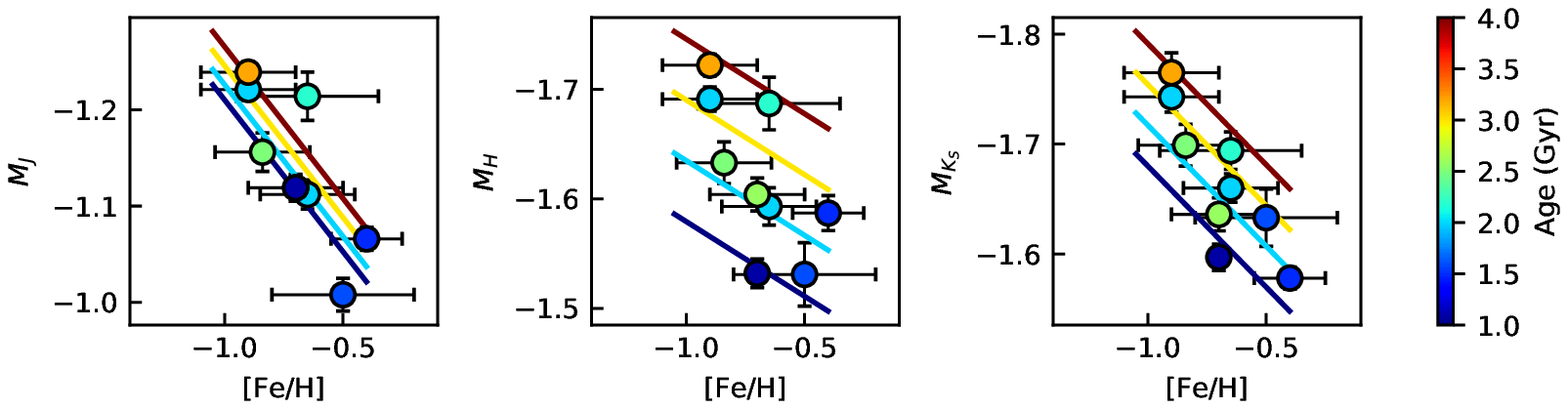}
    \caption{Mean RC magnitude versus metallicity in the $J$- (left), $H$- (centre), and $K_{S}$-band (right). The best-fit relations are also plotted for four ages (from blue to brown, $\log t$ = 9.0, 9.2, 9.4, 9.6). Age difference is shown by colour scales.}
    \label{fig:RC_metal_mag_fit}
\end{figure*}

Figures~\ref{fig:RC_age_mag_fit} and \ref{fig:RC_metal_mag_fit} show the distribution of absolute magnitudes with fitted lines. The population effect of RC stars with the ages of 1.1--3.2~Gyr and the metallicities from $-$0.90 to $-$0.40~dex can be corrected with these relations. As predicted by the model predictions, the regression coefficient for [Fe/H] in $J$-band is larger than those in the $H$- and $K_{S}$-band.
\begin{table*}
\caption{The derived RMSs of differences of absolute magnitudes and colours between observational data and calculated values from rom equations (\ref{eq:Mag_fit}), (\ref{eq:Mag_fit_2}) - (\ref{eq:Mag_fit_4})}
\label{tab:Mag_RMS}
\begin{tabular}{ccccc}
\hline
& Eq. (\ref{eq:Mag_fit}) & Eq. (\ref{eq:Mag_fit_2}) & Eq. (\ref{eq:Mag_fit_3}) & Eq. (\ref{eq:Mag_fit_4})\\
\hline
$J$ & $0.054 \pm 0.019$ & $0.061 \pm 0.021$ & $0.070 \pm 0.024$ & $0.220 \pm 0.071$\\
$H$ & $0.058 \pm 0.021$ & $0.083 \pm 0.029$ & $0.054 \pm 0.019$ & $0.140 \pm 0.045$\\
$K_{S}$ & $0.047 \pm 0.016$ & $0.073 \pm 0.026$ & $0.056 \pm 0.019$ & $0.138 \pm 0.047$\\
\hline
$J - H$ & $0.035 \pm 0.012$ & $0.063 \pm 0.022$ & $0.032 \pm 0.011$ & $0.084 \pm 0.028$\\
$J - K_{S}$ & $0.063 \pm 0.022$ & $0.085 \pm 0.029$ & $0.096 \pm 0.033$ & $0.341 \pm 0.112$\\
$H - K_{S}$ & $0.044 \pm 0.016$ & $0.048 \pm 0.017$ & $0.067 \pm 0.023$ & $0.336 \pm 0.054$\\
\hline
\end{tabular}
\end{table*}

As can be seen in Figure~\ref{fig:Age_and_Metallicity}, the ages of clusters are correlated with the metallicities (the correlation coefficient is $-$0.612 for the nine clusters used for the fitting). Therefore, it is possible that multicollinearity occurs. To check the presence of multicollinearity, we calculated partial correlation coefficients. The partial correlation coefficients between ages and absolute magnitudes are $-$0.247, $-$0.585, and $-$0.540, and the partial correlation coefficients between metallicities and absolute magnitudes are 0.695, 0.384, and 0.706 for $J$-, $H$-, and $K_{S}$-bands, respectively. This means that absolute magnitude depends on both age and metallicity. The correlation coefficients between ages and absolute magnitudes are $-$0.610, $-$0.752, and $-$0.747, and the correlation coefficients between metallicities and absolute magnitudes are 0.809, 0.660, and 0.828 for $J$-, $H$-, and $K_{S}$-bands, respectively. These correlation coefficients have the same signs as the partial correlation coefficients, and the values are comparable. Furthermore, the correlation coefficients have the same sings as the regression coefficients $a$ and $b$. Therefore, multicollinearity does not matter much.

We also fitted the RC colours with the form of equation~(\ref{eq:Mag_fit}), and obtained
\begin{align}
J - H = (0.178 \pm 0.064) \log t + (0.179 \pm 0.053) [\mathrm{Fe}/\mathrm{H}] \notag \\
 -(1.054 \pm 0.577) \label{eq:J_H}\\
J - K_{S} = (0.087 \pm 0.146) \log t + (0.095 \pm 0.120) [\mathrm{Fe}/\mathrm{H}] \notag \\
-(0.212 \pm 1.314) \label{eq:J_K}\\
H - K_{S} = (-0.092 \pm 0.110) \log t -(0.084 \pm 0.090) [\mathrm{Fe}/\mathrm{H}] \notag \\
 + (0.842 \pm 0.990) \label{eq:H_K}.
\end{align}
The adjusted $R^{2}$ values are 0.569, -0.203, -0.143. As is the case of the absolute magnitudes, we conducted leave-one-out cross-validation for the same forms of equations (\ref{eq:Mag_fit}), (\ref{eq:Mag_fit_2}) - (\ref{eq:Mag_fit_4}). The calculated RMSs are presented in Table \ref{tab:Mag_RMS}. The RMSs of colour differences have similar or larger values for the higher order polynomial functions, and thus we selected the form of equation (\ref{eq:Mag_fit}) as the best model following one standard error rule.

\begin{figure*}
    \includegraphics{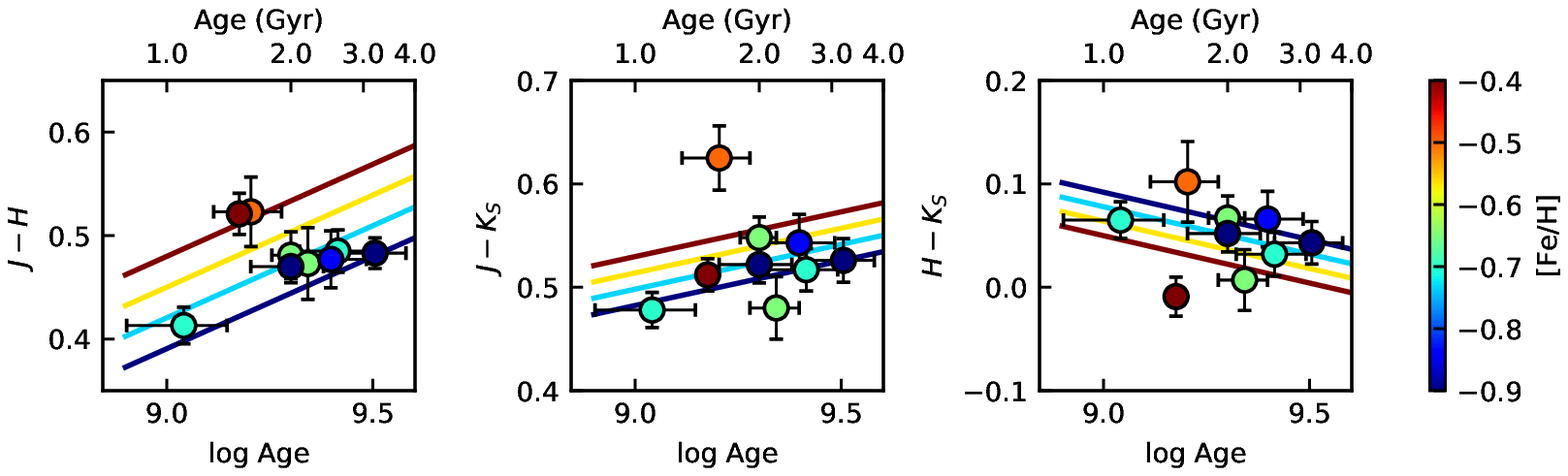}
    \caption{Mean RC colour versus age for the $J - H$ (left), $J - K_{S}$ (centre), and $H - K_{S}$ (right). The best-fit relations are also plotted for four metallicities (from blue to brown, $-$0.90, $-$0.73, $-$0.57, $-$0.40~dex). Metallicity difference is shown by colour scales.}
    \label{fig:RC_color_age_mag_fit}
\end{figure*}

\begin{figure*}
    \includegraphics{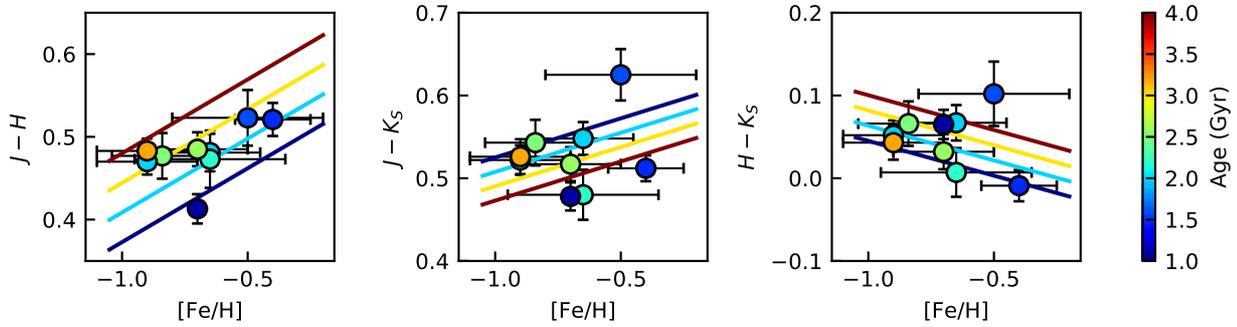}
    \caption{Mean RC colour versus metallicity for the $J - H$ (left), $J - K_{S}$ (centre), and $H - K_{S}$ (right). The best-fit relations are also plotted for four ages (from blue to brown, $\log t$ = 9.0, 9.2, 9.4, 9.6). Age difference is shown by colour scales.}
    \label{fig:RC_color_metal_mag_fit}
\end{figure*}

Figures~\ref{fig:RC_color_age_mag_fit} and \ref{fig:RC_color_metal_mag_fit} show the mean RC colors as a function of age and metallicity with fitted lines. Comparing the absolute magnitudes, the population effect for the RC colours are smaller. In particular, $J - K_{S}$ and $H - K_{S}$ have nearly constant values, and the mean values of nine clusters are $J - K_{S} = 0.528 \pm 0.015$ and $H - K_{S} = 0.047 \pm 0.011$, respectively. These values are used as the intrinsic RC colours at least within the ages of 1.1--3.2~Gyr and the [Fe/H] of $-$0.90 to $-$0.40~dex. The partial correlation coefficients between ages and absolute magnitudes are 0.192 and $-$0.331, and the partial correlation coefficients between metallicities and absolute magnitudes are 0.278 and $-$0.372 for $J - K_{S}$ and $H - K_{S}$, respectively. This also supports that these colours have no strong dependence of absolute magnitude on age and metallicity. These colours can be used as an interstellar extinction probe. The average value of $J - H$ is $0.480 \pm 0.010$, although the population effect is slightly stronger. The partial correlation coefficients between ages and absolute magnitudes are 0.696, and the partial correlation coefficients between metallicities and absolute magnitudes are 0.781 for $J - H$. This means that $J - H$ colour depends on both age and metallicity. For more metal-rich RC stars, \citet{LJP2012} obtained $J - H = 0.506$, $J - K_{S} = 0.629$ and $H - K_{S} = 0.123$ in the solar neighbourhood, and \citet{GRZ2012} derived $J - K_{S} = 0.68$ in Baade's Window. These values are slightly higher than our results. Therefore, attention should be paid for applying these colours to RC stars with near solar metallicities.

During the fitting, the errors on age and [Fe/H] are not considered. The actual errors of the coefficients, $a$, $b$, and $c$, would therefore be larger than those shown in equations (\ref{eq:M_J})~-~(\ref{eq:M_K}) and (\ref{eq:J_H})~-~(\ref{eq:H_K}), because of the large errors on [Fe/H]. In addition, the errors on coefficients are already large because of the small number of samples. It means that, with the current observational results, it is difficult to insist that the coefficients $a$ and $b$ for $M_{J}$, $M_{H}$, $M_{K_{S}}$, and $J - H$ have non-zero values. However, the partial correlation coefficients suggest that $a$, and $b$ for these absolute magnitudes and $J - H$ colour have nonzero values. Deeper NIR observations of the LMC clusters and careful metallicity measurements of them are necessary to determine the coefficients more accurately.

\section{Conclusions}\label{sec:Conclusions}
In this paper, we investigate the age and metallicity dependence on the RC magnitudes, $m_{J}$, $m_{H}$, and $m_{K_{S}}$, and their colours $J - H$, $J - K_{S}$, and $H - K_{S}$. Most of our samples consist of the clusters with young age and low metallicity. The age and metallicity are different from clusters in the Milky Way. We obtained the relation to correct the population effect for the absolute magnitudes and colours of RC stars with the age of 1.1--3.2~Gyr and [Fe/H] from -0.90 to -0.40~dex, although the errors on the coefficients are large. We confirmed that the population effect for $J - K_{S}$ and $H - K_{S}$ colours is very small within these age and metallicity ranges. When we use RC stars as a standard candle, we can accurately estimate interstellar extinction without suffering from the population effect by using these colours. In model comparison, our observational data show good agreement with the prediction by \citet{G2016} between 1.6--3.2~Gyr. The averaged $M_{K_{S}}$ is consistent with previous work, but $M_{J}$ and $M_{K_{S}}$ are slightly bright. This discrepancy may be caused by the population effect.

\section*{Acknowledgements}
This work was supported by Grant-in-Aid for Japan Society for the Promotion of Science Research Fellow Grant Number JP16J00865. The IRSF project is a collaboration between Nagoya University and the SAAO supported by the Grants-in-Aid for Scientific Research on Priority Areas (A) (No. 10147207 and No. 10147214) and Optical \& Near-Infrared Astronomy Inter-University Cooperation Program, from the Ministry of Education, Culture, Sports, Science and Technology (MEXT) of Japan and the National Research Foundation (NRF) of South Africa. This publication makes use of data products from the Two Micron All Sky Survey, which is a joint project of the University of Massachusetts and the Infrared Processing and Analysis Center/California Institute of Technology, funded by the National Aeronautics and Space Administration and the National Science Foundation.




\bibliographystyle{mnras}
\bibliography{reference} 





\bsp	
\label{lastpage}
\end{document}